\documentclass[10pt,draft,onecolumn]{IEEEtran}
\usepackage{amsmath,amssymb,threeparttable,placeins,makecell,stfloats,cite}
\usepackage[dvips]{graphicx,epstopdf}
\usepackage[usenames]{color}
\usepackage{amsfonts}
\usepackage{latexsym}
\usepackage{subfigure}
\usepackage[justification=centering]{caption}
\usepackage{floatrow}
\newtheorem{theorem}{{{\textit{Theorem}}}}

\newtheorem{lemma}{{{\textit{Lemma}}}}
\newtheorem{corollary}{{{{\textit{Corollary}}}}}

\newtheorem{definition}{{{\textit{Definition}}}}

\newtheorem{example}{{{\textit{Example}}}}

\newtheorem{construction}{{{\textit{Construction}}}}

\begin{document}
	
	\title{A Generalized Construction of Multiple Complete Complementary Codes and Asymptotically Optimal Aperiodic Quasi-Complementary Sequence Sets}
	\author{ Zhengchun Zhou,
		Fangrui Liu,
		Avik Ranjan Adhikary,
		and Pingzhi Fan.
		\thanks{
			%

			Zhengchun Zhou, Fangrui Liu, and Avik Ranjan Adhikary are with  School of Mathematics, Southwest Jiaotong University,
			Chengdu, 610031, China, and also with the State Key Laboratory of Cryptology,
			Beijing 100878, China. E-mail:{\tt zzc@swjtu.edu.cn, 754521389@qq.com, Avik.Adhikary@ieee.org} .
			Pingzhi Fan is with Institute of Mobile Communications, Southwest Jiaotong University,
			Chengdu, 610031, China. E-mail: {\tt p.fan@ieee.org}.}

	}
	\maketitle
	
%
%
%

\begin{abstract}
In recent years, complete complementary codes (CCCs) and quasi-complementary sequence sets (QCSSs) have found many important applications in multi-carrier code-division multiple-access (MC-CDMA) systems for their good correlation properties. In this paper, we propose a generic construction of multiple sets of CCCs over $\mathbb{Z}_N$, consisting of sequences of length $N$, where $N\geq 3$ is an arbitrary odd integer. Interestingly, the maximum inter-set aperiodic cross-correlation magnitude of the proposed CCCs is upper bounded by $N$. It turns out that the combination of the generated CCCs results in a new set of sequences to obtain asymptotically optimal and near-optimal aperiodic QCSSs. The proposed construction includes a recent optimal construction of QCSSs with prime length as a special case and leads to asymptotically optimal QCSSs with new flexible parameters.
\end{abstract}


\section{INTRODUCTION}
\IEEEPARstart{P}{erfect} complementary sequence set (PCSS) refers to a set of  two-dimensional matrices with non-trivial autocorrelations and cross-correlations summing to zero for any non-zero time-shift. Such sequence sets have wide applications due to their ideal correlation properties, such as reducing the peak-to-average power ratio (PAPR) \cite{Davis-1999}, channel estimation \cite{Spasojevic-2001}, \cite{Wang-2007}, RADAR waveform design \cite{Pezeshki-2008}, etc. In typical applications of complementary set, each constituent
sequence should be sent out in a separate channel and
thus each complementary set is transmitted in a “multichannel” system. When a multi-carrier code division multiple access (MC-CDMA)
transmission of complementary sets is considered,
due to the mutual orthogonality of the complementary sets of the PCSS, a PCSS-MC-CDMA
with $M$ sub-carriers is capable of supporting at most $M$ users \cite{zilongccc,chen1}.


In search of sequence sets which can support a large number of users in MC-CDMA systems, Liu \textit{et al.} \cite{zilong111} initially proposed low correlation zone complementary sequence sets (LCZ-CSSs). Further Liu \textit{et al.} \cite{zilong13} generalized the concept and proposed quasi-complementary sequence sets (QCSSs) in 2013, which also include Z-complementary sequence sets (ZCSSs) \cite{zhou15,zilong14,yubo14,Avik2,sarkar18,Avik5,chen17}. A $(K,M,N,\delta_{\max })$- QCSS is a set of $K$ $M\times N$ sequence sets, where $K$ denotes the set size (i.e., the number of users), $M$ denotes the flock size (i.e. the number of sub-carriers), $N$ denotes the sequence length, and $\delta_{\max }$ is the maximum periodic or aperiodic correlation tolerance, depending on what we are considering. When $\delta_{\max}=0$ and $K\leq M$, QCSS becomes PCSS. A PCSS is called a complete complementary code (CCC) when $K=M$ \cite{zilong13}.

In 1974, Welch \cite{welch} derived a lower bound for the correlation of sequences. Later in 2011, Liu \textit{et al.} \cite{zilong11,zilong14_1} tightened the lower bound for aperiodic QCSSs when the set size $K\geq 3M$. Liu \textit{et al.} \cite{zilong17} further tightened the lower bound for the cases of aperiodic QCSSs, when $\frac{\lfloor\frac{\pi^2M}{4}\rfloor}{M}<\frac{K}{M}<3+\frac{1}{M}$ for sufficiently large $N$.

In \cite{zilong13}, Liu \textit{et al.} stated the construction of aperiodic QCSSs, achieving the bounds proposed in \cite{zilong14_1}, as an open problem. Since then a lot of research have been going on, both for periodic and aperiodic cases, for the construction of QCSSs with various parameters and correlation properties. However, most of the works, including \cite{zilong13}, are related to periodic QCSS. Constructions of asymptotically optimal and near-optimal periodic QCSS based on difference sets and almost difference sets were reported in \cite{Li18} and \cite{Li19_1}, respectively. In \cite{Li19} and \cite{Li19_2}, Li \textit{et al.} gave a construction of periodic QCSSs based on characters over finite fields. To the best of the authors' knowledge, very few works are available in the literature on the constructions of asymptotically optimal aperiodic QCSSs. Working on this problem, recently in 2019 Li \textit{et al.} \cite{Li19_3} proposed asymptotically optimal aperiodic QCSSs based on low-correlation CSSs. In another work of Li \textit{et al.} \cite{Li19_4}, the authors proposed asymptotically optimal aperiodic QCSSs based on multiple sets of CCCs consisting sequences of prime length\footnote{Although the authors reported the QCSS to be near-optimal, the QCSS are actually asymptotically optimal with respect to a tighter correlation bound.}.


The constructions of aperiodic QCSS reported in \cite{Li19_3} and \cite{Li19_4} are all based on functions from $\mathbb{Z}_N$ to itself, where $N$ is restricted to primes or power of primes. The parameters of the asymptotically optimal  QCSSs are listed in Table \ref{tab_intro}.
\begin{table}
	\small
	\resizebox{\textwidth}{!}{
	\caption{Parameters of asymptotically optimal aperiodic QCSSs.\label{tab_intro}}
	\begin{tabular}{|c|c|c|c|c|c|c|}
		\hline
		Ref. & $K$ & $M$ &  $N$ &$\delta_{\max }$ &  Alphabet&  Parameter constraint(s)   \\\hline
\cite{Li19_4} & $p(p-1)$ & $p$ & $p$& $p$ & $\mathbb{Z}_p$ &  $p$ is an odd prime. \\
		\hline
		Th. 1 \cite{Li19_3} & $q(q+1)$ &  $q$ & $q$&$q$ &$\mathbb{Z}_q$ & $q$ is power of a prime. \\
		\hline
		Th. 3 \cite{Li19_3} & $q^2$ & $q$ & $q-1$&$q$ & $\mathbb{Z}_q$ & $q\geq 5$ is power of prime. \\
		\hline
		Proposed & $N(p_0-1)$ & $N$ & $N$&$N$ & $\mathbb{Z}_N$ &  \makecell{$N\geq 5$ is odd, $N=p_0^{e_0}p_1^{e_1}\dots p_{n-1}^{e_{n-1}}$,\\where $p_0<p_1<\cdots<p_{n-1}$ \\ are prime factors of $N$.} \\
		\hline
	\end{tabular} }\\
\end{table}
In search of aperiodic QCSS with more generalized parameters, we propose a new permutation on $\mathbb{Z}_N$, where $N$ is an arbitrary odd integer. Based on this new permutation, we generate multiple sets of $(N,N,N)$- CCCs. Combining the proposed CCCs, we construct ($N(p_0-1),N,N,N$)- QCSS, where $p_0$ is the least prime factor of $N$. Since for $p_0\geq 5$, our proposed QCSS satisfies the condition $K\geq 3M,~M\geq2$ and $N\geq 2$, we consider the correlation bound proposed by Liu \textit{et al.} \cite{zilong14_1} to check the optimality condition. Interestingly, all the proposed CCCs have an inter-set cross correlation magnitude, upper bounded by $N$. Hence, all the proposed QCSSs are asymptotically optimal, having the optimality factor $\rho$ approximately equal to $1$. For the cases when $N\geq3$ has minimum prime factor $p_0=3$, we calculate the optimality factor $\rho$, with the help of the Welch bound \cite{welch}. The proposed QCSSs are near-optimal when $p_0=3$, with the value of optimality factor $\rho$ approaching towards $2$. Our construction is more generalized as compared to \cite{Li19_3} and \cite{Li19_4}, since $N>2$ can also be any odd integer other than a prime or power of prime. In particular, the construction in \cite{Li19_4} can be seen as a special case of our construction when $N$ is a prime.


The rest of this paper is organized as follows. In Section II, we introduce some correlation bounds of QCSS. In Section III, we propose a new permutation, and construct multiple sets of CCCs based on the new permutation. In Section IV, we obtain asymptotically optimal aperiodic QCSS by combining these CCCs into a new set. In Section V, we make a comparison of our work with the existing works in the literature. We conclude our work in Section VI.

\section{Bounds of QCSS}

In this section we recall some correlation bounds of QCSS. Before that let us fix some notations which will be used throughout the paper.
\begin{itemize}
	\item  $N\geq 3$ is an odd integer such that $N=p_0^{e_0}p_1^{e_1}\dots p_{n-1}^{e_{n-1}}$, where $p_0<p_1<\cdots<p_{n-1}$ are prime factors of $N$ and $e_0,e_1,\cdots,e_{n-1}$ are non-negative integers.
	\item $\mathbb{Z}_N$ denotes the ring of integers modulo $N$.
	\item $\omega_{N}=e^{\frac{2\pi i}{N}}$ is a primitive $N$-th root of unity.
	\item $\mathbb{Z}_N^*=\{a:1\leq a<N,~\gcd(a,N)=1\}$.
	\item $\mathbb{Z}_N^{**}=\mathbb{Z}^*_N\setminus\{1\}$.
	\item $\phi \circ \psi$ denotes the composition of functions $\phi$ and $\psi$, i.e., $\phi \circ \psi(x)=\phi(\psi(x))$.
	\item $\mathfrak{C}$ denotes a set of two dimensional matrices of sequence sets.
	\item $\mathcal{C}$ denotes a two dimensional matrix of sequences.
	\item $C$ denotes a sequence or a row of a two dimensional matrix of sequences.
\end{itemize}

\begin{definition}
Let $u=\left(u_{0}, u_{1}, \cdots, u_{N-1}\right)$ and $v=\left(v_{0}, v_{1}, \cdots, v_{N-1}\right)$ denote two complex-valued sequences with length $N$. The aperiodic correlation function between $u$ and $v$ is defined as
\begin{equation}
\tilde{R}_{u, v}(\tau)=\left\{\begin{array}{ll}{\sum_{t=0}^{N-1-\tau} u_{t} v_{t+\tau}^{*},} & {0 \leq \tau \leq N-1} \\ {\sum_{t=0}^{N-1-\tau} u_{t-\tau} v_{t}^{*},} & {-N+1 \leq \tau \leq 0}\end{array}\right.
\end{equation}
where $x^*$ denotes the complex conjugation of a complex-valued sequence $x$.
\end{definition}

\begin{definition}
	Let $\mathfrak{C}=\left\{\mathcal{C}^{0}, \mathcal{C}^{1}, \cdots, \mathcal{C}^{K-1}\right\}$ be a set of $K$ two-dimensional matrices of sequence sets, each of size $M\times N$, i.e.,
	\begin{equation}\label{eq2}
\mathcal{C}^{k}=\left[ \begin{array}{c}{C_{0}^{k}} \\ {C_{1}^{k}} \\ {\vdots} \\ {C_{M-1}^{k}}\end{array}\right]_{M \times N}, 0 \leq k \leq K-1,
	\end{equation}
	where the $m$-th row $ C_{m}^{k}=\left(c_{m, 0}^{k}, c_{m, 1}^{k}, \cdots, c_{m, N-1}^{k}\right)$ denotes the $m$-th  length-$N$ constituent sequence, $0 \leq m \leq M-1$. The set $\mathfrak{C}$ is called a quasi-complementary sequence set (QCSS) if for any $\mathcal{C}^{k_1},\mathcal{C}^{k_2}\in \mathfrak{C}$, $0\leq k_1,k_2
	\leq K-1$,
	$0 \leq \tau \leq N-1, k_{1} \neq k_{2}$ or $0<\tau \leq N-1, k_{1}=k_{2}$,
	\begin{equation}
	|\tilde{R}_{\mathcal{C}^{k_1},\mathcal{C}^{k_2}}(\tau)|=\left|\sum_{m=0}^{M-1} \tilde{R}_{C_{m}^{k_{1}},
		{C_{m}^{k_{2}}}}(\tau)\right| \leq \delta_{\max },
	\end{equation}
	where
	$K$ is the set size, $M$ is the size of each sequence set, $N$ is the length of constituent sequences, and $\delta_{\max }$ is the maximum aperiodic cross-correlation magnitude  of $\mathfrak{C}$.  For simplicity, the set $\mathfrak{C}$ is  denoted by $(K,M,N, \delta_{\max })$-QCSS. Specially, we call $\mathfrak{C}$ a $(M,M,N,0)\equiv (M,M,N)$-CCC if $K = M$ and $\delta_{\max }=0$.
\end{definition}


The following lemma gives a lower bound of $\delta_{\max }$.
\begin{lemma} {\cite{welch}}\label{lem1}
	For an aperiodic $(K,M,N, \delta_{\max })$-QCSS, the parameters satisfy:
	\begin{equation}
	\delta_{\max } \geq M N \cdot \sqrt{\frac{\left(\frac{K}{M}-1\right)}{K(2 N-1)-1}}.
	\end{equation}
\end{lemma}

For $K\geq 3M$, $M\geq2$ and $N \geq 2$, Liu \textit{et al.} \cite{zilong14_1} proposed a tighter correlation bound for aperiodic QCSS, which is given in the following lemma.

\begin{lemma} {\cite{zilong14_1}}
	For an aperiodic $(K,M,N, \delta_{\max })$-QCSS with $K\geq 3M$, $M\geq2$ and $N \geq 2$, we have
	\begin{equation}\label{eq-1}
	\delta_{\max } \geq \sqrt{MN\left(1-2\sqrt{\frac{M}{3K}}\right)}
	\end{equation}
\end{lemma}

In this paper, when $K\geq 3M$, i.e., for $N\geq 5$ having the minimum prime factor $p_0$ of $N$ greater than $3$, a QCSS is optimal if $\delta_{\max }$ achieves the lower bound in (\ref{eq-1}). Therefore, for these cases we define the optimality factor $\rho$ as follows
\begin{equation}
\rho=\frac{\delta_{\text {max}}}{\sqrt{MN\left(1-2\sqrt{\frac{M}{3K}}\right)}}.
\end{equation}
For the case when $K\ngeq 3M$, i.e., when $N\geq3$ has minimum prime factor $p_0=3$, the optimality factor $\rho$ is defined as follows
\begin{equation}
\rho=\frac{\delta_{\text {max}}}{M N \cdot \sqrt{\frac{\left(\frac{K}{M}-1\right)}{K(2 N-1)-1}}}.
\end{equation}
Note that $\rho\geq 1$. A QCSS is optimal when $\rho=1$ and near-optimal if $1<\rho\leq2$.

\section{Multiple CCCs With Low Maximum Inter-set Aperiodic Cross-Correlation Magnitude}
In this section, first we derive a new permutation over $\mathbb{Z}_N$. Then  we shall present a construction of multiple CCCs with low maximum inter-set aperiodic cross-correlation magnitude, based on that permutation. The main idea behind the construction is inspired by the recent works in \cite{Li19_4} and \cite{zhou18}. We extend the constructions of \cite{Li19_4} by generalizing $N$ to any odd integer. It will be seen that the construction can lead to new multiple CCCs with low maximum inter-set aperiodic correlation magnitude. Before that, we recall the following lemma.	
	\begin{lemma}\cite{zhou18}\label{lem_zhou}
		Let $\xi(x)=x^e$, where $e$ is a positive integer with gcd$(p-1,e)=1$. Then $\xi$ is a permutation on $\mathbb{Z}_p$, where $p$ is any odd prime, such that the equation $\xi(x+a)\equiv c\xi(x) \pmod p$ has only one solution in $\mathbb{Z}_p$ for any $a\in \mathbb{Z}_p$ and any $c\in \mathbb{Z}^{**}_p$.
	\end{lemma}

Let $N\geq 3$ be an odd integer of the form $N=p_0^{e_0}p_1^{e_1}\dots p_{n-1}^{e_{n-1}}$, where $p_0<p_1<\cdots<p_{n-1}$ are prime factors of $N$ and $e_0,e_1,\cdots,e_{n-1}$ are non-negative integers. For an integer $i\in \mathbb{Z}_N$, let us define one-to-one mappings $\psi$ and $\phi$ given in (\ref{new_permu}) and (\ref{rev_new_permu}), respectively.
\begin{figure*}
	\begin{flushleft}
		\begin{equation}\label{new_permu}
		\psi:\mathbb{Z}_N\rightarrow \underbrace{\mathbb{Z}_{p_0}\times \cdots \times \mathbb{Z}_{p_0}}_{e_0 \text{ times}}\times \underbrace{\mathbb{Z}_{p_1}\times \cdots \times \mathbb{Z}_{p_1}}_{e_1 \text{ times}}\times \cdots \times \underbrace{\mathbb{Z}_{p_{n-1}}\times \cdots \times \mathbb{Z}_{p_{n-1}}}_{e_{n-1} \text{ times}}
		\end{equation}
		defined by
		\begin{equation}
		\psi(i)=(i_{(0,p_0)},\dots,i_{(e_0-1,p_0)},i_{(0,p_1)},\dots,i_{(e_1-1,p_1)},\dots, i_{(0,p_{n-1})},\dots,i_{(e_{n-1}-1,p_{n-1})}) \text{ where } i_{(\cdot,p_j)}\in \mathbb{Z}_{p_j}
		\end{equation}
		and
		\begin{equation}
		\begin{split}
		&i=p_0^{e_0-1}p_1^{e_1}\dots p_{n-1}^{e_{n-1}}i_{(0,p_0)}+p_0^{e_0-2}p_1^{e_1}\dots p_{n-1}^{e_{n-1}}i_{(1,p_0)}+\cdots+ p_1^{e_1}\dots p_{n-1}^{e_{n-1}}i_{(e_0-1,p_0)}+\\&\hspace{1cm} p_1^{e_1-1}p_2^{e_2}\dots p_{n-1}^{e_{n-1}}i_{(0,p_1)}+\cdots+p_2^{e_2}\dots p_{n-1}^{e_{n-1}}i_{(e_1-1,p_1)} + \dots +p_{n-1}^{e_{n-1}-1}i_{(0,p_{n-1})}+\cdots+i_{(e_{n-1}-1,p_{n-1})}.
		\end{split}
		\end{equation}
	\end{flushleft}
\end{figure*}
\begin{figure*}
	\begin{flushleft}
		\begin{equation}\label{rev_new_permu}
		\phi: \underbrace{\mathbb{Z}_{p_0}\times \cdots \times \mathbb{Z}_{p_0}}_{e_0 \text{ times}}\times \underbrace{\mathbb{Z}_{p_1}\times \cdots \times \mathbb{Z}_{p_1}}_{e_1 \text{ times}}\times \cdots \times \underbrace{\mathbb{Z}_{p_{n-1}}\times \cdots \times \mathbb{Z}_{p_{n-1}}}_{e_{n-1} \text{ times}}\rightarrow\mathbb{Z}_N,
		\end{equation}
		defined by
		\begin{equation}\label{neq12}
		\begin{split}
		&\phi(\psi(i))=
		p_0^{e_0-1}p_1^{e_1}\dots p_{n-1}^{e_{n-1}}i_{(0,p_0)}+p_0^{e_0-2}p_1^{e_1}\dots p_{n-1}^{e_{n-1}}i_{(1,p_0)}+\cdots+ p_1^{e_1}\dots p_{n-1}^{e_{n-1}}i_{(e_0-1,p_0)}+\\&\hspace{1cm} p_1^{e_1-1}p_2^{e_2}\dots p_{n-1}^{e_{n-1}}i_{(0,p_1)}+\cdots+p_2^{e_2}\dots p_{n-1}^{e_{n-1}}i_{(e_1-1,p_1)} + \dots +p_{n-1}^{e_{n-1}-1}i_{(0,p_{n-1})}+\cdots+\xi(i_{(e_{n-1}-1,p_{n-1})}),
		\\&\text{where }\xi \text{ is a permutation on }\mathbb{Z}_{p_{n-1}} \text{ as given in Lemma \ref{lem_zhou}}.
		\end{split}
		\end{equation}
	\end{flushleft}
\end{figure*}
Then the permutation $\pi:\mathbb{Z}_N\rightarrow \mathbb{Z}_N$ is given as $\pi=\phi \circ \psi$, i.e., $\pi(x)=\phi\circ\psi(x)=\phi(\psi(x))$, where $\phi(\psi(x))$ is defined in (\ref{neq12}).

To illustrate the permutation $\pi$ on $\mathbb{Z}_{N}$, let us have the following example.
\begin{example}
	Let us consider $N=15$. Here $p_0=3$ and $p_1=5$. So any element $i$ in $Z_{15}$ can be written as $i=5i_0+i_1$, where $i_0\in \mathbb{Z}_3$ and $i_1 \in \mathbb{Z}_5$. According to Lemma \ref{lem_zhou}, let $\xi:\mathbb{Z}_5 \rightarrow \mathbb{Z}_5$ be defined as $\xi(x)=x^3 \pmod 5$. Then $\psi:\mathbb{Z}_{15}\rightarrow \mathbb{Z}_3 \times \mathbb{Z}_5$ is given by $\psi(x)=(i_0,~i_1)$ where $x=p_1i_0+i_1=5i_0+i_1$. $\phi:\mathbb{Z}_3 \times \mathbb{Z}_5 \rightarrow \mathbb{Z}_{15}$ is given by $\phi(i_0,i_1)=p_1i_0+\xi(i_1)=5i_0+\xi(i_1)$. Therefore, the permutation $\pi:\mathbb{Z}_{15} \rightarrow \mathbb{Z}_{15}$ is given by $\pi(x)=\phi \circ \psi (x)$. $\mathbb{Z}_{15}=\{0,1,2,3,4,5,6,7,8,9,10,11,12,13,14\}$ and $\pi(\mathbb{Z}_{15})=\{0,1,3,2,4,5,6,8,7,9,10,11,13,12,14\}$.
\end{example}

\begin{lemma}\label{new_per_lem}
	Let $\pi(x)=\phi \circ \psi(x)$ be a permutation on $\mathbb{Z}_N$, defined as above. Then the equation $\pi(x+\tau)\equiv c\pi(x)\pmod N$ has only one solution in $\mathbb{Z}_N$ for any $\tau\in \mathbb{Z}_N$ and any $c\in \mathbb{Z}_{p_0}^{**}$.
\end{lemma}

\begin{IEEEproof}
	Let $\tau\in \mathbb{Z}_N$, then $\tau$ can be written as
	\begin{equation}
	\begin{split}
	&\tau=p_0^{e_0-1}p_1^{e_1}\dots p_{n-1}^{e_{n-1}}\tau_{(0,p_0)}+p_0^{e_0-2}p_1^{e_1}\dots p_{n-1}^{e_{n-1}}\tau_{(1,p_0)}+\\&\hspace{0.5cm}\cdots+ p_1^{e_1}\dots p_{n-1}^{e_{n-1}}\tau_{(e_0-1,p_0)}+p_1^{e_1-1}p_2^{e_2}\dots p_{n-1}^{e_{n-1}}\tau_{(0,p_1)}+\\&\hspace{0.5cm} \cdots+p_2^{e_2}\dots p_{n-1}^{e_{n-1}}\tau_{(e_1-1,p_1)} + \dots +p_{n-1}^{e_{n-1}-1}\tau_{(0,p_{n-1})}+\\&\hspace{0.5cm} \cdots+\tau_{(e_{n-1}-1,p_{n-1})},
	\end{split}
	\end{equation}
	where $\tau_{(\cdot,p_{j})}\in \mathbb{Z}_{p_j}$. Also let
	\begin{equation}
	\delta_{i_{(\cdot,p_{j})},\tau_{(\cdot,p_{j})}}=\begin{cases}
	0, &\text{ if }i_{(\cdot,p_{j})}+\tau_{(\cdot,p_{j})}<p_j\\
	1, &\text{ if }i_{(\cdot,p_{j})}+\tau_{(\cdot,p_{j})}\geq p_j\\
	\end{cases}
	\end{equation}
	Now, $\pi(x)$ is given by
	\begin{equation}\label{eq15}
	\begin{split}
		&\pi(x)=\\&p_0^{e_0-1}p_1^{e_1}\dots p_{n-1}^{e_{n-1}}i_{(0,p_0)}+p_0^{e_0-2}p_1^{e_1}\dots p_{n-1}^{e_{n-1}}i_{(1,p_0)}+\\&\cdots+ p_1^{e_1}\dots p_{n-1}^{e_{n-1}}i_{(e_0-1,p_0)}+p_1^{e_1-1}p_2^{e_2}\dots p_{n-1}^{e_{n-1}}i_{(0,p_1)}+\\& \cdots+p_2^{e_2}\dots p_{n-1}^{e_{n-1}}i_{(e_1-1,p_1)} + \dots +p_{n-1}^{e_{n-1}-1}i_{(0,p_{n-1})}+\\&\cdots+\xi(i_{(e_{n-1}-1,p_{n-1})}),
	\end{split}
	\end{equation}
	and $\pi(x+\tau)$ is given in (\ref{eq16}).
	\begin{figure*}
		\begin{equation}\label{eq16}
		\begin{split}
		\pi(x+\tau)=\\&p_0^{e_0-1}p_1^{e_1}\dots p_{n-1}^{e_{n-1}}(i_{(0,p_0)}+\tau_{(0,p_0)}+\delta_{i_{(1,p_{0})},\tau_{(1,p_{0})}}\pmod {p_0})+\\&p_0^{e_0-2}p_1^{e_1}\dots p_{n-1}^{e_{n-1}}(i_{(1,p_0)}+\tau_{(1,p_0)}+\delta_{i_{(2,p_{0})},\tau_{(2,p_{0})}}\pmod {p_0})+\cdots\\&+ p_1^{e_1}\dots p_{n-1}^{e_{n-1}}(i_{(e_0-1,p_0)}+\tau_{(e_0-1,p_0)}+\delta_{i_{(0,p_{1})},\tau_{(0,p_{1})}}\pmod {p_0})+\\& p_1^{e_1-1}p_2^{e_2}\dots p_{n-1}^{e_{n-1}}(i_{(0,p_1)}+\tau_{(0,p_1)}+\delta_{i_{(1,p_{1})},\tau_{(1,p_{1})}}\pmod {p_1})+\cdots\\&+p_2^{e_2}\dots p_{n-1}^{e_{n-1}}(i_{(e_1-1,p_1)}+\tau_{(e_1-1,p_1)}+\delta_{i_{(0,p_{2})},\tau_{(0,p_{2})}}\pmod {p_1}) + \dots\\& +p_{n-1}^{e_{n-1}-1}(i_{(0,p_{n-1})}+\tau_{(0,p_{n-1})}+\delta_{i_{(1,p_{n-1})},\tau_{(1,p_{n-1})}}\pmod {p_{n-1}})\\&+\cdots+\xi(i_{(e_{n-1}-1,p_{n-1})}+\tau_{(e_{n-1}-1,p_{n-1})}\pmod {p_{n-1}}).
		\end{split}
		\end{equation}
	\end{figure*}

	From Lemma \ref{lem_zhou},
	\begin{equation}
	\begin{split}
	\xi(i_{(e_{n-1}-1,p_{n-1})}+&\tau_{(e_{n-1}-1,p_{n-1})})\equiv\\& c\xi(i_{(e_{n-1}-1,p_{n-1})}) \pmod {p_{n-1}}
	\end{split}
	\end{equation}
	has a unique solution say $i^\prime_{(e_{n-1}-1,p_{n-1})}\in \mathbb{Z}_{p_{n-1}}$. From the equation $\pi(x+\tau)\equiv c\pi(x)\pmod N$, (\ref{eq15}) and (\ref{eq16}), we get unique $i_{(m,p_j)}^\prime$ given by
	\begin{equation}
	i_{(m,p_j)}^\prime=\frac{\tau_{(m,p_j)}+\delta_{i_{(m+1,p_{j})},\tau_{(m+1,p_{j})}}}{c-1} \pmod {p_j},
	\end{equation}
	where $0\leq m<e_j$ and $0\leq j<n$.

%
Therefore, for this unique $i^\prime_{(e_{n-1}-1,p_{n-1})}$ and $i_{(m,p_j)}^\prime$'s we get a unique $x$ in $\mathbb{Z}_N$ for which the statement of the theorem holds, when $c\in \mathbb{Z}_{p_0}^{**}$.	  			
\end{IEEEproof}

Using the permutation defined above, satisfying Lemma \ref{new_per_lem}, we generate multiple sets of CCCs in the following construction.

\begin{construction}\label{new_const}
		Let $N\geq 3$ be an odd positive integer such that $N=p_0^{e_0}p_1^{e_1}\dots p_{n-1}^{e_{n-1}}$, where $p_0<p_1<\cdots<p_{n-1}$ are prime factors of $N$, and $e_0,e_1,\cdots,e_{n-1}$ are non-negative integers. Also, let $\pi$ be the permutation over $\mathbb{Z}_N$, defined above satisfying Lemma \ref{new_per_lem}. Then for each $k\in \mathbb{Z}^*_{p_0}$, $m\in  \mathbb{Z}_N$, and $s\in \mathbb{Z}_N$, define a function $f_{s}^{(k,m)}$ from $\mathbb{Z}_N$ to itself, where
	\begin{equation}\label{eq_new_const2}
	f_{s}^{(k,m)}(t)=
	ks\cdot \pi(t)+m t \pmod {N}.
	\end{equation}
		For each $k\in \mathbb{Z}^*_{p_0}$, define a set
	\begin{equation}\label{C^k_2}
	\mathcal{\mathfrak{C}}^{k}=\{\mathcal{C}^{(k,0)},\mathcal{C}^{(k,1)},\cdots,
	\mathcal{C}^{(k,N-1)}\},
	\end{equation}
	where
	\[
	\mathcal{C}^{(k,m)}\hspace{-0.1cm}=\hspace{-0.1cm}
	\begin{bmatrix}
	C_{0}^{(k,m)}\\
	C_{1}^{(k,m)}\\
	\vdots  \\
	C_{N-1}^{(k,m)} \\
	\end{bmatrix}\hspace{-0.1cm}
	=\hspace{-0.1cm}
	\begin{bmatrix}
	C_{0,0}^{(k,m)} , C_{0,1}^{(k,m)} , \cdots , C_{0,N-1}^{(k,m)} \\
	C_{1,0}^{(k,m)} , C_{1,1}^{(k,m)} , \cdots , C_{1,N-1}^{(k,m)} \\
	\vdots  \\
	C_{N-1,0}^{(k,m)} , C_{N-1,1}^{(k,m)} , \cdots , C_{N-1,N-1}^{(k,m)} \\
	\end{bmatrix}
	\]
	and
	$$
	C_{s,t}^{(k,m)}=\omega_{N}^{f_{s}^{(k,m)}(t)} \textrm{~for~each~}0\leq t\leq N-1.
	$$
\end{construction}

For the sequence sets generated by Construction \ref{new_const}, we have the following result.
\begin{theorem}\label{theorem2}
	Let $\mathfrak{C}^{k},~1\leq k <p_0$, be the multiple sequence sets generated by Construction \ref{new_const} using the function $f_{s}^{(k,m)}: \mathbb{Z}_{N}\rightarrow\mathbb{Z}_{N}$, as defined in (\ref{eq_new_const2}). Then,
	\begin{enumerate}
		\item Each sequence set $\mathfrak{C}^{k}$ is a $(N,N,N)$-CCC.
		\item The inter-set cross-correlation between any two distinct CCCs ${\mathfrak{C}}^{k_1}$ and ${\mathfrak{C}}^{k_2}$ is upper bounded by $N$, where $k_1\neq k_2$. That is,
		
		\begin{equation}
			|\sum\limits_{s=0}^{N-1} \tilde{R}_{C_{s}^{\left(k_{1}, m_{1}\right)},C_{s}^{\left(k_{2}, m_{2}\right)} }(\tau)| \leq N,
		\end{equation} 
		for all $0\leq\tau \leq N-1, k_1\neq k_2 \in \mathbb{Z}^*_{p_0}$ and $0\leq m_1,m_2\leq N-1.$
	\end{enumerate}
\end{theorem}

\begin{IEEEproof}
	Please see Appendix A.
\end{IEEEproof}

\begin{example}\label{ex11}
	Let $N=35$ with $p_0=5$ and $p_1=7$. Define $\xi:\mathbb{Z}_7\rightarrow \mathbb{Z}_7$ as $\xi(x)=x^5$, a permutation over $\mathbb{Z}_7$. Then $\pi=\phi \circ \psi (x)=7i_0+\xi(i_1)$. By Theorem \ref{theorem2}, we get four $(35,35,35)$-CCCs, $\mathfrak{C}^{1}, \mathfrak{C}^{2}, \mathfrak{C}^{3}, \mathfrak{C}^{4}$. The first complementary set of first two CCCs are shown in Table \ref{tab_ex2}.
\end{example}
\begin{table*}
	\caption{The first complementary set of $\mathfrak{C}^1$ and $\mathfrak{C}^2$ of Example \ref{ex11}.\label{tab_ex2}}
		\small
	\resizebox{0.9\textwidth}{!}{
	\begin{tabular}{|l *{33}l l|}
		\hline
		\multicolumn{35}{|c|}{\Large$\mathcal{C}^{(1,0)}$}\\ \hline
		0 & 0 & 0 & 0 & 0 & 0 & 0 & 0 & 0 & 0 & 0 & 0 & 0 & 0 & 0 & 0 & 0 & 0 & 0 & 0 & 0 & 0 & 0 & 0 & 0 & 0 & 0 & 0 & 0 & 0 & 0 & 0 & 0 & 0 & 0 \\
		0 & 1 & 4 & 5 & 2 & 3 & 6 & 7 & 8 & 11 & 12 & 9 & 10 & 13 & 14 & 15 & 18 & 19 & 16 & 17 & 20 & 21 & 22 & 25 & 26 & 23 & 24 & 27 & 28 & 29 & 32 & 33 & 30 & 31 & 34 \\
		0 & 2 & 8 & 10 & 4 & 6 & 12 & 14 & 16 & 22 & 24 & 18 & 20 & 26 & 28 & 30 & 1 & 3 & 32 & 34 & 5 & 7 & 9 & 15 & 17 & 11 & 13 & 19 & 21 & 23 & 29 & 31 & 25 & 27 & 33 \\
		0 & 3 & 12 & 15 & 6 & 9 & 18 & 21 & 24 & 33 & 1 & 27 & 30 & 4 & 7 & 10 & 19 & 22 & 13 & 16 & 25 & 28 & 31 & 5 & 8 & 34 & 2 & 11 & 14 & 17 & 26 & 29 & 20 & 23 & 32 \\
		0 & 4 & 16 & 20 & 8 & 12 & 24 & 28 & 32 & 9 & 13 & 1 & 5 & 17 & 21 & 25 & 2 & 6 & 29 & 33 & 10 & 14 & 18 & 30 & 34 & 22 & 26 & 3 & 7 & 11 & 23 & 27 & 15 & 19 & 31 \\
		0 & 5 & 20 & 25 & 10 & 15 & 30 & 0 & 5 & 20 & 25 & 10 & 15 & 30 & 0 & 5 & 20 & 25 & 10 & 15 & 30 & 0 & 5 & 20 & 25 & 10 & 15 & 30 & 0 & 5 & 20 & 25 & 10 & 15 & 30 \\
		0 & 6 & 24 & 30 & 12 & 18 & 1 & 7 & 13 & 31 & 2 & 19 & 25 & 8 & 14 & 20 & 3 & 9 & 26 & 32 & 15 & 21 & 27 & 10 & 16 & 33 & 4 & 22 & 28 & 34 & 17 & 23 & 5 & 11 & 29 \\
		0 & 7 & 28 & 0 & 14 & 21 & 7 & 14 & 21 & 7 & 14 & 28 & 0 & 21 & 28 & 0 & 21 & 28 & 7 & 14 & 0 & 7 & 14 & 0 & 7 & 21 & 28 & 14 & 21 & 28 & 14 & 21 & 0 & 7 & 28 \\
		0 & 8 & 32 & 5 & 16 & 24 & 13 & 21 & 29 & 18 & 26 & 2 & 10 & 34 & 7 & 15 & 4 & 12 & 23 & 31 & 20 & 28 & 1 & 25 & 33 & 9 & 17 & 6 & 14 & 22 & 11 & 19 & 30 & 3 & 27 \\
		0 & 9 & 1 & 10 & 18 & 27 & 19 & 28 & 2 & 29 & 3 & 11 & 20 & 12 & 21 & 30 & 22 & 31 & 4 & 13 & 5 & 14 & 23 & 15 & 24 & 32 & 6 & 33 & 7 & 16 & 8 & 17 & 25 & 34 & 26 \\
		0 & 10 & 5 & 15 & 20 & 30 & 25 & 0 & 10 & 5 & 15 & 20 & 30 & 25 & 0 & 10 & 5 & 15 & 20 & 30 & 25 & 0 & 10 & 5 & 15 & 20 & 30 & 25 & 0 & 10 & 5 & 15 & 20 & 30 & 25 \\
		0 & 11 & 9 & 20 & 22 & 33 & 31 & 7 & 18 & 16 & 27 & 29 & 5 & 3 & 14 & 25 & 23 & 34 & 1 & 12 & 10 & 21 & 32 & 30 & 6 & 8 & 19 & 17 & 28 & 4 & 2 & 13 & 15 & 26 & 24 \\
		0 & 12 & 13 & 25 & 24 & 1 & 2 & 14 & 26 & 27 & 4 & 3 & 15 & 16 & 28 & 5 & 6 & 18 & 17 & 29 & 30 & 7 & 19 & 20 & 32 & 31 & 8 & 9 & 21 & 33 & 34 & 11 & 10 & 22 & 23 \\
		0 & 13 & 17 & 30 & 26 & 4 & 8 & 21 & 34 & 3 & 16 & 12 & 25 & 29 & 7 & 20 & 24 & 2 & 33 & 11 & 15 & 28 & 6 & 10 & 23 & 19 & 32 & 1 & 14 & 27 & 31 & 9 & 5 & 18 & 22 \\
		0 & 14 & 21 & 0 & 28 & 7 & 14 & 28 & 7 & 14 & 28 & 21 & 0 & 7 & 21 & 0 & 7 & 21 & 14 & 28 & 0 & 14 & 28 & 0 & 14 & 7 & 21 & 28 & 7 & 21 & 28 & 7 & 0 & 14 & 21 \\
		0 & 15 & 25 & 5 & 30 & 10 & 20 & 0 & 15 & 25 & 5 & 30 & 10 & 20 & 0 & 15 & 25 & 5 & 30 & 10 & 20 & 0 & 15 & 25 & 5 & 30 & 10 & 20 & 0 & 15 & 25 & 5 & 30 & 10 & 20 \\
		0 & 16 & 29 & 10 & 32 & 13 & 26 & 7 & 23 & 1 & 17 & 4 & 20 & 33 & 14 & 30 & 8 & 24 & 11 & 27 & 5 & 21 & 2 & 15 & 31 & 18 & 34 & 12 & 28 & 9 & 22 & 3 & 25 & 6 & 19 \\
		0 & 17 & 33 & 15 & 34 & 16 & 32 & 14 & 31 & 12 & 29 & 13 & 30 & 11 & 28 & 10 & 26 & 8 & 27 & 9 & 25 & 7 & 24 & 5 & 22 & 6 & 23 & 4 & 21 & 3 & 19 & 1 & 20 & 2 & 18 \\
		0 & 18 & 2 & 20 & 1 & 19 & 3 & 21 & 4 & 23 & 6 & 22 & 5 & 24 & 7 & 25 & 9 & 27 & 8 & 26 & 10 & 28 & 11 & 30 & 13 & 29 & 12 & 31 & 14 & 32 & 16 & 34 & 15 & 33 & 17 \\
		0 & 19 & 6 & 25 & 3 & 22 & 9 & 28 & 12 & 34 & 18 & 31 & 15 & 2 & 21 & 5 & 27 & 11 & 24 & 8 & 30 & 14 & 33 & 20 & 4 & 17 & 1 & 23 & 7 & 26 & 13 & 32 & 10 & 29 & 16 \\
		0 & 20 & 10 & 30 & 5 & 25 & 15 & 0 & 20 & 10 & 30 & 5 & 25 & 15 & 0 & 20 & 10 & 30 & 5 & 25 & 15 & 0 & 20 & 10 & 30 & 5 & 25 & 15 & 0 & 20 & 10 & 30 & 5 & 25 & 15 \\
		0 & 21 & 14 & 0 & 7 & 28 & 21 & 7 & 28 & 21 & 7 & 14 & 0 & 28 & 14 & 0 & 28 & 14 & 21 & 7 & 0 & 21 & 7 & 0 & 21 & 28 & 14 & 7 & 28 & 14 & 7 & 28 & 0 & 21 & 14 \\
		0 & 22 & 18 & 5 & 9 & 31 & 27 & 14 & 1 & 32 & 19 & 23 & 10 & 6 & 28 & 15 & 11 & 33 & 2 & 24 & 20 & 7 & 29 & 25 & 12 & 16 & 3 & 34 & 21 & 8 & 4 & 26 & 30 & 17 & 13 \\
		0 & 23 & 22 & 10 & 11 & 34 & 33 & 21 & 9 & 8 & 31 & 32 & 20 & 19 & 7 & 30 & 29 & 17 & 18 & 6 & 5 & 28 & 16 & 15 & 3 & 4 & 27 & 26 & 14 & 2 & 1 & 24 & 25 & 13 & 12 \\
		0 & 24 & 26 & 15 & 13 & 2 & 4 & 28 & 17 & 19 & 8 & 6 & 30 & 32 & 21 & 10 & 12 & 1 & 34 & 23 & 25 & 14 & 3 & 5 & 29 & 27 & 16 & 18 & 7 & 31 & 33 & 22 & 20 & 9 & 11 \\
		0 & 25 & 30 & 20 & 15 & 5 & 10 & 0 & 25 & 30 & 20 & 15 & 5 & 10 & 0 & 25 & 30 & 20 & 15 & 5 & 10 & 0 & 25 & 30 & 20 & 15 & 5 & 10 & 0 & 25 & 30 & 20 & 15 & 5 & 10 \\
		0 & 26 & 34 & 25 & 17 & 8 & 16 & 7 & 33 & 6 & 32 & 24 & 15 & 23 & 14 & 5 & 13 & 4 & 31 & 22 & 30 & 21 & 12 & 20 & 11 & 3 & 29 & 2 & 28 & 19 & 27 & 18 & 10 & 1 & 9 \\
		0 & 27 & 3 & 30 & 19 & 11 & 22 & 14 & 6 & 17 & 9 & 33 & 25 & 1 & 28 & 20 & 31 & 23 & 12 & 4 & 15 & 7 & 34 & 10 & 2 & 26 & 18 & 29 & 21 & 13 & 24 & 16 & 5 & 32 & 8 \\
		0 & 28 & 7 & 0 & 21 & 14 & 28 & 21 & 14 & 28 & 21 & 7 & 0 & 14 & 7 & 0 & 14 & 7 & 28 & 21 & 0 & 28 & 21 & 0 & 28 & 14 & 7 & 21 & 14 & 7 & 21 & 14 & 0 & 28 & 7 \\
		0 & 29 & 11 & 5 & 23 & 17 & 34 & 28 & 22 & 4 & 33 & 16 & 10 & 27 & 21 & 15 & 32 & 26 & 9 & 3 & 20 & 14 & 8 & 25 & 19 & 2 & 31 & 13 & 7 & 1 & 18 & 12 & 30 & 24 & 6 \\
		0 & 30 & 15 & 10 & 25 & 20 & 5 & 0 & 30 & 15 & 10 & 25 & 20 & 5 & 0 & 30 & 15 & 10 & 25 & 20 & 5 & 0 & 30 & 15 & 10 & 25 & 20 & 5 & 0 & 30 & 15 & 10 & 25 & 20 & 5 \\
		0 & 31 & 19 & 15 & 27 & 23 & 11 & 7 & 3 & 26 & 22 & 34 & 30 & 18 & 14 & 10 & 33 & 29 & 6 & 2 & 25 & 21 & 17 & 5 & 1 & 13 & 9 & 32 & 28 & 24 & 12 & 8 & 20 & 16 & 4 \\
		0 & 32 & 23 & 20 & 29 & 26 & 17 & 14 & 11 & 2 & 34 & 8 & 5 & 31 & 28 & 25 & 16 & 13 & 22 & 19 & 10 & 7 & 4 & 30 & 27 & 1 & 33 & 24 & 21 & 18 & 9 & 6 & 15 & 12 & 3 \\
		0 & 33 & 27 & 25 & 31 & 29 & 23 & 21 & 19 & 13 & 11 & 17 & 15 & 9 & 7 & 5 & 34 & 32 & 3 & 1 & 30 & 28 & 26 & 20 & 18 & 24 & 22 & 16 & 14 & 12 & 6 & 4 & 10 & 8 & 2 \\
		0 & 34 & 31 & 30 & 33 & 32 & 29 & 28 & 27 & 24 & 23 & 26 & 25 & 22 & 21 & 20 & 17 & 16 & 19 & 18 & 15 & 14 & 13 & 10 & 9 & 12 & 11 & 8 & 7 & 6 & 3 & 2 & 5 & 4 & 1 \\  \hline
		\multicolumn{35}{|c|}{\Large$\mathcal{C}^{(2,0)}$}\\ \hline	
	0 & 0 & 0 & 0 & 0 & 0 & 0 & 0 & 0 & 0 & 0 & 0 & 0 & 0 & 0 & 0 & 0 & 0 & 0 & 0 & 0 & 0 & 0 & 0 & 0 & 0 & 0 & 0 & 0 & 0 & 0 & 0 & 0 & 0 & 0 \\
	0 & 2 & 8 & 10 & 4 & 6 & 12 & 14 & 16 & 22 & 24 & 18 & 20 & 26 & 28 & 30 & 1 & 3 & 32 & 34 & 5 & 7 & 9 & 15 & 17 & 11 & 13 & 19 & 21 & 23 & 29 & 31 & 25 & 27 & 33 \\
	0 & 4 & 16 & 20 & 8 & 12 & 24 & 28 & 32 & 9 & 13 & 1 & 5 & 17 & 21 & 25 & 2 & 6 & 29 & 33 & 10 & 14 & 18 & 30 & 34 & 22 & 26 & 3 & 7 & 11 & 23 & 27 & 15 & 19 & 31 \\
	0 & 6 & 24 & 30 & 12 & 18 & 1 & 7 & 13 & 31 & 2 & 19 & 25 & 8 & 14 & 20 & 3 & 9 & 26 & 32 & 15 & 21 & 27 & 10 & 16 & 33 & 4 & 22 & 28 & 34 & 17 & 23 & 5 & 11 & 29 \\
	0 & 8 & 32 & 5 & 16 & 24 & 13 & 21 & 29 & 18 & 26 & 2 & 10 & 34 & 7 & 15 & 4 & 12 & 23 & 31 & 20 & 28 & 1 & 25 & 33 & 9 & 17 & 6 & 14 & 22 & 11 & 19 & 30 & 3 & 27 \\
	0 & 10 & 5 & 15 & 20 & 30 & 25 & 0 & 10 & 5 & 15 & 20 & 30 & 25 & 0 & 10 & 5 & 15 & 20 & 30 & 25 & 0 & 10 & 5 & 15 & 20 & 30 & 25 & 0 & 10 & 5 & 15 & 20 & 30 & 25 \\
	0 & 12 & 13 & 25 & 24 & 1 & 2 & 14 & 26 & 27 & 4 & 3 & 15 & 16 & 28 & 5 & 6 & 18 & 17 & 29 & 30 & 7 & 19 & 20 & 32 & 31 & 8 & 9 & 21 & 33 & 34 & 11 & 10 & 22 & 23 \\
	0 & 14 & 21 & 0 & 28 & 7 & 14 & 28 & 7 & 14 & 28 & 21 & 0 & 7 & 21 & 0 & 7 & 21 & 14 & 28 & 0 & 14 & 28 & 0 & 14 & 7 & 21 & 28 & 7 & 21 & 28 & 7 & 0 & 14 & 21 \\
	0 & 16 & 29 & 10 & 32 & 13 & 26 & 7 & 23 & 1 & 17 & 4 & 20 & 33 & 14 & 30 & 8 & 24 & 11 & 27 & 5 & 21 & 2 & 15 & 31 & 18 & 34 & 12 & 28 & 9 & 22 & 3 & 25 & 6 & 19 \\
	0 & 18 & 2 & 20 & 1 & 19 & 3 & 21 & 4 & 23 & 6 & 22 & 5 & 24 & 7 & 25 & 9 & 27 & 8 & 26 & 10 & 28 & 11 & 30 & 13 & 29 & 12 & 31 & 14 & 32 & 16 & 34 & 15 & 33 & 17 \\
	0 & 20 & 10 & 30 & 5 & 25 & 15 & 0 & 20 & 10 & 30 & 5 & 25 & 15 & 0 & 20 & 10 & 30 & 5 & 25 & 15 & 0 & 20 & 10 & 30 & 5 & 25 & 15 & 0 & 20 & 10 & 30 & 5 & 25 & 15 \\
	0 & 22 & 18 & 5 & 9 & 31 & 27 & 14 & 1 & 32 & 19 & 23 & 10 & 6 & 28 & 15 & 11 & 33 & 2 & 24 & 20 & 7 & 29 & 25 & 12 & 16 & 3 & 34 & 21 & 8 & 4 & 26 & 30 & 17 & 13 \\
	0 & 24 & 26 & 15 & 13 & 2 & 4 & 28 & 17 & 19 & 8 & 6 & 30 & 32 & 21 & 10 & 12 & 1 & 34 & 23 & 25 & 14 & 3 & 5 & 29 & 27 & 16 & 18 & 7 & 31 & 33 & 22 & 20 & 9 & 11 \\
	0 & 26 & 34 & 25 & 17 & 8 & 16 & 7 & 33 & 6 & 32 & 24 & 15 & 23 & 14 & 5 & 13 & 4 & 31 & 22 & 30 & 21 & 12 & 20 & 11 & 3 & 29 & 2 & 28 & 19 & 27 & 18 & 10 & 1 & 9 \\
	0 & 28 & 7 & 0 & 21 & 14 & 28 & 21 & 14 & 28 & 21 & 7 & 0 & 14 & 7 & 0 & 14 & 7 & 28 & 21 & 0 & 28 & 21 & 0 & 28 & 14 & 7 & 21 & 14 & 7 & 21 & 14 & 0 & 28 & 7 \\
	0 & 30 & 15 & 10 & 25 & 20 & 5 & 0 & 30 & 15 & 10 & 25 & 20 & 5 & 0 & 30 & 15 & 10 & 25 & 20 & 5 & 0 & 30 & 15 & 10 & 25 & 20 & 5 & 0 & 30 & 15 & 10 & 25 & 20 & 5 \\
	0 & 32 & 23 & 20 & 29 & 26 & 17 & 14 & 11 & 2 & 34 & 8 & 5 & 31 & 28 & 25 & 16 & 13 & 22 & 19 & 10 & 7 & 4 & 30 & 27 & 1 & 33 & 24 & 21 & 18 & 9 & 6 & 15 & 12 & 3 \\
	0 & 34 & 31 & 30 & 33 & 32 & 29 & 28 & 27 & 24 & 23 & 26 & 25 & 22 & 21 & 20 & 17 & 16 & 19 & 18 & 15 & 14 & 13 & 10 & 9 & 12 & 11 & 8 & 7 & 6 & 3 & 2 & 5 & 4 & 1 \\
	0 & 1 & 4 & 5 & 2 & 3 & 6 & 7 & 8 & 11 & 12 & 9 & 10 & 13 & 14 & 15 & 18 & 19 & 16 & 17 & 20 & 21 & 22 & 25 & 26 & 23 & 24 & 27 & 28 & 29 & 32 & 33 & 30 & 31 & 34 \\
	0 & 3 & 12 & 15 & 6 & 9 & 18 & 21 & 24 & 33 & 1 & 27 & 30 & 4 & 7 & 10 & 19 & 22 & 13 & 16 & 25 & 28 & 31 & 5 & 8 & 34 & 2 & 11 & 14 & 17 & 26 & 29 & 20 & 23 & 32 \\
	0 & 5 & 20 & 25 & 10 & 15 & 30 & 0 & 5 & 20 & 25 & 10 & 15 & 30 & 0 & 5 & 20 & 25 & 10 & 15 & 30 & 0 & 5 & 20 & 25 & 10 & 15 & 30 & 0 & 5 & 20 & 25 & 10 & 15 & 30 \\
	0 & 7 & 28 & 0 & 14 & 21 & 7 & 14 & 21 & 7 & 14 & 28 & 0 & 21 & 28 & 0 & 21 & 28 & 7 & 14 & 0 & 7 & 14 & 0 & 7 & 21 & 28 & 14 & 21 & 28 & 14 & 21 & 0 & 7 & 28 \\
	0 & 9 & 1 & 10 & 18 & 27 & 19 & 28 & 2 & 29 & 3 & 11 & 20 & 12 & 21 & 30 & 22 & 31 & 4 & 13 & 5 & 14 & 23 & 15 & 24 & 32 & 6 & 33 & 7 & 16 & 8 & 17 & 25 & 34 & 26 \\
	0 & 11 & 9 & 20 & 22 & 33 & 31 & 7 & 18 & 16 & 27 & 29 & 5 & 3 & 14 & 25 & 23 & 34 & 1 & 12 & 10 & 21 & 32 & 30 & 6 & 8 & 19 & 17 & 28 & 4 & 2 & 13 & 15 & 26 & 24 \\
	0 & 13 & 17 & 30 & 26 & 4 & 8 & 21 & 34 & 3 & 16 & 12 & 25 & 29 & 7 & 20 & 24 & 2 & 33 & 11 & 15 & 28 & 6 & 10 & 23 & 19 & 32 & 1 & 14 & 27 & 31 & 9 & 5 & 18 & 22 \\
	0 & 15 & 25 & 5 & 30 & 10 & 20 & 0 & 15 & 25 & 5 & 30 & 10 & 20 & 0 & 15 & 25 & 5 & 30 & 10 & 20 & 0 & 15 & 25 & 5 & 30 & 10 & 20 & 0 & 15 & 25 & 5 & 30 & 10 & 20 \\
	0 & 17 & 33 & 15 & 34 & 16 & 32 & 14 & 31 & 12 & 29 & 13 & 30 & 11 & 28 & 10 & 26 & 8 & 27 & 9 & 25 & 7 & 24 & 5 & 22 & 6 & 23 & 4 & 21 & 3 & 19 & 1 & 20 & 2 & 18 \\
	0 & 19 & 6 & 25 & 3 & 22 & 9 & 28 & 12 & 34 & 18 & 31 & 15 & 2 & 21 & 5 & 27 & 11 & 24 & 8 & 30 & 14 & 33 & 20 & 4 & 17 & 1 & 23 & 7 & 26 & 13 & 32 & 10 & 29 & 16 \\
	0 & 21 & 14 & 0 & 7 & 28 & 21 & 7 & 28 & 21 & 7 & 14 & 0 & 28 & 14 & 0 & 28 & 14 & 21 & 7 & 0 & 21 & 7 & 0 & 21 & 28 & 14 & 7 & 28 & 14 & 7 & 28 & 0 & 21 & 14 \\
	0 & 23 & 22 & 10 & 11 & 34 & 33 & 21 & 9 & 8 & 31 & 32 & 20 & 19 & 7 & 30 & 29 & 17 & 18 & 6 & 5 & 28 & 16 & 15 & 3 & 4 & 27 & 26 & 14 & 2 & 1 & 24 & 25 & 13 & 12 \\
	0 & 25 & 30 & 20 & 15 & 5 & 10 & 0 & 25 & 30 & 20 & 15 & 5 & 10 & 0 & 25 & 30 & 20 & 15 & 5 & 10 & 0 & 25 & 30 & 20 & 15 & 5 & 10 & 0 & 25 & 30 & 20 & 15 & 5 & 10 \\
	0 & 27 & 3 & 30 & 19 & 11 & 22 & 14 & 6 & 17 & 9 & 33 & 25 & 1 & 28 & 20 & 31 & 23 & 12 & 4 & 15 & 7 & 34 & 10 & 2 & 26 & 18 & 29 & 21 & 13 & 24 & 16 & 5 & 32 & 8 \\
	0 & 29 & 11 & 5 & 23 & 17 & 34 & 28 & 22 & 4 & 33 & 16 & 10 & 27 & 21 & 15 & 32 & 26 & 9 & 3 & 20 & 14 & 8 & 25 & 19 & 2 & 31 & 13 & 7 & 1 & 18 & 12 & 30 & 24 & 6 \\
	0 & 31 & 19 & 15 & 27 & 23 & 11 & 7 & 3 & 26 & 22 & 34 & 30 & 18 & 14 & 10 & 33 & 29 & 6 & 2 & 25 & 21 & 17 & 5 & 1 & 13 & 9 & 32 & 28 & 24 & 12 & 8 & 20 & 16 & 4 \\
	0 & 33 & 27 & 25 & 31 & 29 & 23 & 21 & 19 & 13 & 11 & 17 & 15 & 9 & 7 & 5 & 34 & 32 & 3 & 1 & 30 & 28 & 26 & 20 & 18 & 24 & 22 & 16 & 14 & 12 & 6 & 4 & 10 & 8 & 2 \\   \hline
	\end{tabular}}
\end{table*}

In the following section, we propose the following two constructions:
\begin{enumerate}
	\item For $N\geq 5$ having minimum prime factor $p_0>3$, we propose a construction of asymptotically optimal aperiodic QCSSs.
	\item For $N\geq 3$ having minimum prime factor $p_0=3$, we propose a construction of asymptotically near-optimal aperiodic QCSSs.
\end{enumerate}

\section{Proposed Construction of Asymptotically Optimal QCSSs}
In this section, from multiple CCCs with low maximum aperiodic cross-correlation magnitude, we will obtain an aperiodic asymptotically optimal and near-optimal QCSS by combining these complementary sequence sets together, as shown in the following theorem and corollary.
%
%

\begin{figure}
	\includegraphics[draft=false,width=\textwidth]{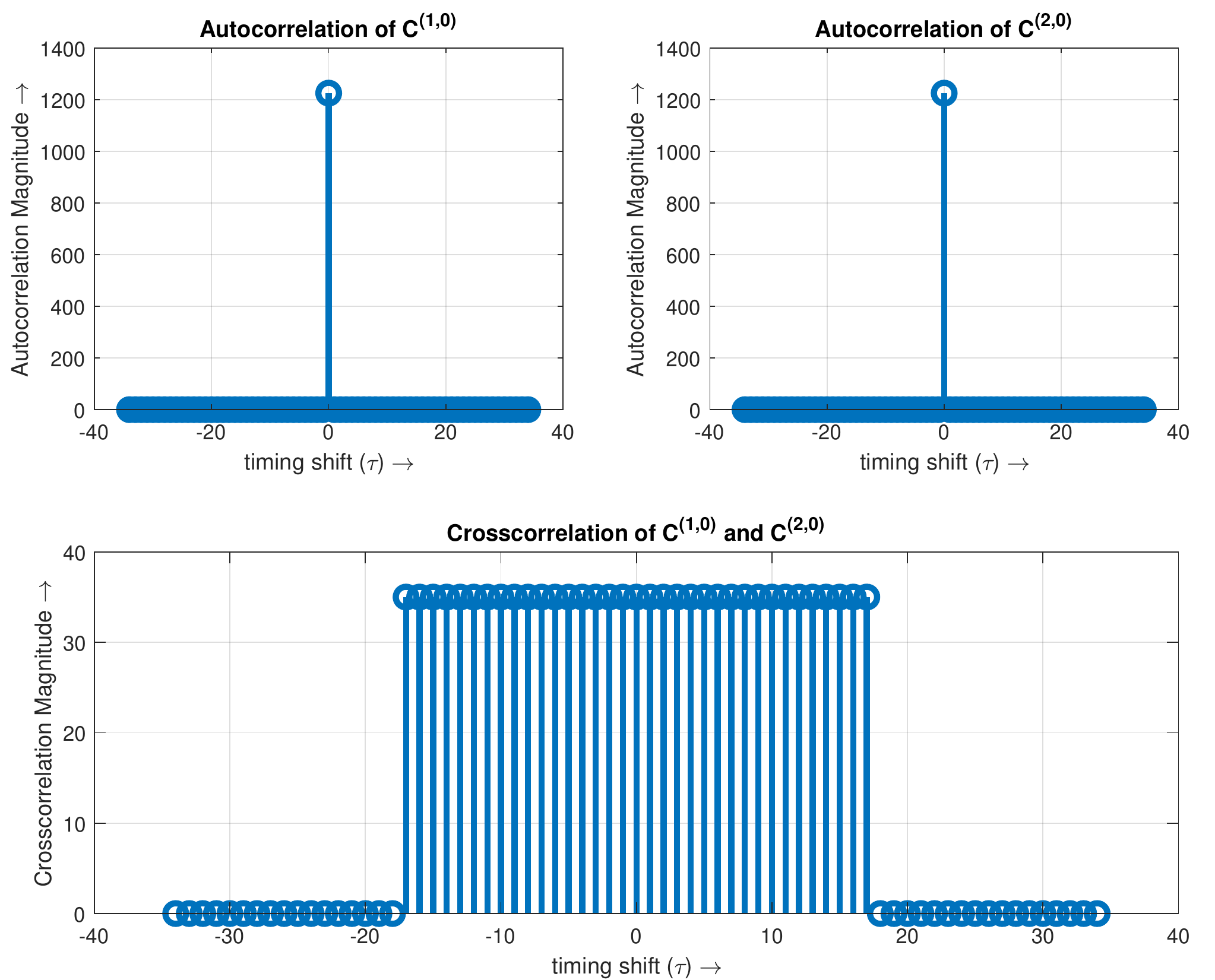}
	\caption{A glimpse of correlation magnitudes of the QCSS in \textit{Example \ref{ex11}}.}\label{label-a}
\end{figure}

\begin{theorem}\label{cor_n_5}
Let $N\geq 5$ be an odd positive integer such that $N=p_0^{e_0}p_1^{e_1}\dots p_{n-1}^{e_{n-1}}$, where $p_0<p_1<\cdots<p_{n-1}$ are prime factors of $N$ with $p_0\geq 5$ and $e_0,e_1,\cdots,e_{n-1}$ are non-negative integers. Also, let $\mathfrak{C}^{k}\text{ for }1\leq k <p_0$ be generated according to Theorem \ref{theorem2} and
$\mathfrak{C}=\mathfrak{C}^{1} \cup \mathfrak{C}^{2} \cup \cdots \cup \mathfrak{C}^{p_0-1}$, then the sequence set $\mathfrak{C}$ is an asymptotically optimal aperiodic $(N(p_0-1),N,N,N)$-QCSS.
\end{theorem}

\begin{IEEEproof}
	According to Theorem \ref{theorem2}, each sequence set $\mathfrak{C}^{k}$ is a $(N,N,N)$-CCC for each $k\in \mathbb{Z}^*_{p_0}$, and the maximum aperiodic cross-correlation amplitude is $N$. Then $\mathfrak{C}$ is an aperiodic $(K,M,N,\delta_{\max })$-QCSS, where $K=N(p_0-1), ~M=N,~ N=N,~\delta_{\max }=max\{0,N\}=N.$ \par
	
	The optimality factor of $(N(p_0-1),N,N,N)$-QCSS is
	
	\begin{equation}\label{neq21}
		 \rho =\frac{N}{\sqrt{N^{2} (1-2\sqrt{\frac{N}{3N(p_0-1)}} )}}.
	\end{equation}
Taking limit on both sides of (\ref{neq21}) when $p_0\rightarrow +\infty$, we get
\begin{equation}
\begin{aligned}\lim_{p_0\rightarrow +\infty} \rho &=\lim_{p_0\rightarrow +\infty}\frac{N}{\sqrt{N^{2} (1-2\sqrt{\frac{N}{3N(p_0-1)}} )}} \\ &=\lim_{p_0\rightarrow +\infty}\frac{1}{\sqrt{1-\frac{2}{\sqrt{3(p_0-1)}}}}\\&=1.\\
\end{aligned}
\end{equation}
Hence, $\mathfrak{C}$ is an asymptotically optimal aperiodic QCSS.
\end{IEEEproof}

Based on Theorem \ref{cor_n_5}, a list of parameters of asymptotically optimal aperiodic QCSSs are given in Table \ref{tab_composite}.

\begin{table}[h!]
	\small
	\begin{tabular}{|c|c|c|c|c|c|}
	\hline
Alphabet & $K$ & $M$ & $N$ & $\rho$   \\
\hline
$\mathbb{Z}_{5*7}$ & 140 & 35 & 35 & 1.5382  \\
\hline
$\mathbb{Z}_{7*11}$ & 462 & 77 & 77 & 1.3754  \\
\hline
$\mathbb{Z}_{11*13}$ & 1430 & 143 & 143 & 1.2551  \\
\hline
$\mathbb{Z}_{13*17}$ & 2652 & 221 & 221 & 1.2247  \\
\hline
$\mathbb{Z}_{17*19}$ & 5168 & 323 & 323 & 1.1857  \\
\hline
$\mathbb{Z}_{19*23}$ & 7866 & 437 & 437 & 1.1722  \\
\hline
$\mathbb{Z}_{23*31}$ & 15686 & 713 & 713 & 1.1518  \\
\hline
$\mathbb{Z}_{31*37}$ & 34410 & 1147 & 1147 & 1.1257  \\
\hline
$\mathbb{Z}_{37*41}$ & 54612 & 1517 & 1517 & 1.1128  \\
\hline
$\mathbb{Z}_{41*43}$ & 70520 & 1763 & 1763 & 1.1061  \\
\hline
$\mathbb{Z}_{43*47}$ & 84882 & 2021 & 2021 & 1.1031  \\
\hline
$\mathbb{Z}_{53*59}$ & 162604 & 3127 & 3127 & 1.0912  \\
\hline
$\mathbb{Z}_{61*67}$ & 245220 & 4087 & 4087 & 1.0841  \\
\hline
$\mathbb{Z}_{67*71}$ & 313962 & 4757 & 4757 & 1.0797  \\
\hline
$\mathbb{Z}_{71*73}$ & 362810 & 5183 & 5183 & 1.0771  \\
\hline
$\mathbb{Z}_{73*79}$ & 415224 & 5767 & 5767 & 1.0759  \\
\hline
$\mathbb{Z}_{79*83}$ & 511446 & 6557 & 6557 & 1.0726  \\
\hline
$\mathbb{Z}_{83*89}$ & 605734 & 7387 & 7387 & 1.0706  \\
\hline
$\mathbb{Z}_{89*97}$ & 759704 & 8633 & 8633 & 1.0679  \\
\hline
	\end{tabular}\\
	\caption{The parameters of asymptotically optimal aperiodic QCSSs.\label{tab_composite}}
\end{table}

\begin{corollary}\label{case_for_3}
Let $N\geq 3$ be an odd positive integer such that $N=p_0^{e_0}p_1^{e_1}\dots p_{n-1}^{e_{n-1}}$, where $p_0<p_1<\cdots<p_{n-1}$ are prime factors of $N$, $p_0=3$ and $e_0,e_1,\cdots,e_{n-1}$ are non-negative integers. Also, let $\mathfrak{C}^{k}\text{ for }1\leq k <p_0$ be generated according to Theorem \ref{theorem2} and
$\mathfrak{C}=\mathfrak{C}^{1} \cup \mathfrak{C}^{2} \cup \cdots \cup \mathfrak{C}^{p_0-1}$, then the sequence set $\mathfrak{C}$ is a near-optimal aperiodic $(2N,N,N,N)$-QCSS.	
\end{corollary}

\begin{IEEEproof}
	The proof is similar to Theorem \ref{cor_n_5}. However, in this case we will use the Welch bound, given in Lemma \ref{lem1}, since $K \ngeq 3M$.
	Therefore, the optimality factor of $(2N,N,N,N)$-QCSS is
	
	\begin{equation}\label{neq23}
		\rho =\frac{N}{N^2\sqrt{\frac{(\frac{2N}{N}-1)}{2N(2N-1)-1}}}.
	\end{equation}
	Taking limit on both sides of (\ref{neq23}) when $p_0\rightarrow +\infty$, we get
	\begin{equation}
	\begin{aligned} \lim_{N\rightarrow +\infty}\rho &=\lim_{N\rightarrow +\infty}\frac{N}{N^2\sqrt{\frac{(\frac{2N}{N}-1)}{2N(2N-1)-1}}} \\ &=\lim_{N\rightarrow +\infty}\frac{1}{\sqrt{\frac{N^2}{4N^2-2N-1}}}\\&=\lim_{N\rightarrow +\infty} \sqrt{4-\frac{2}{N}-\frac{1}{N^2}}\\&=2.\\
	\end{aligned}
	\end{equation}
\end{IEEEproof}

Based on Corollary \ref{case_for_3}, a list of parameters of asymptotically near-optimal aperiodic QCSSs are given in Table \ref{new_table_for_3}.

\begin{table}[t]
	\small
	\begin{tabular}{|c|c|c|c|c|c|}
		\hline
		Alphabet & $K$ & $M$ & $N$ & $\rho$   \\
		\hline
		$\mathbb{Z}_{3*5}$ & 30 & 15 & 15 & 1.9653 \\
		\hline
		$\mathbb{Z}_{3*7}$ & 42 & 21 & 21 & 1.9755  \\
		\hline
		$\mathbb{Z}_{3*11}$ & 66 & 33 & 33 & 1.9846  \\
		\hline
		$\mathbb{Z}_{3*5*7}$ & 210 & 105 & 105 & 1.9952  \\
		\hline
		$\mathbb{Z}_{3*5*11}$ & 330 & 165 & 165 & 1.9970  \\
		\hline
		$\mathbb{Z}_{3*5*7*11}$ & 2310 & 1155 & 1155 & 1.9996  \\
		\hline
		$\mathbb{Z}_{3*5*7*11*13}$ & 30030 & 15015 & 15015 & 2.0000  \\
		\hline
		$\mathbb{Z}_{3*5*7*11*13*17}$ & 510510 & 255255 & 255255 & 2.0000  \\
		\hline
	\end{tabular}\\
	\caption{The parameters of  near optimal aperiodic QCSSs.\label{new_table_for_3}}
\end{table}

In the following section, we compare our proposed constructions with the previous works.

\section{Comparison With Previous Works}
We compare the optimality factor for the asymptotically optimal QCSS with the results reported in \cite{Li19_3}. We have taken the minimum of all the optimality factors we can get from the previous works reported in \cite{Li19_3}, for each alphabet size, whenever it is possible, and denoted it by $\rho_{\min\_prev}$ for the comparison. $K_{prev}$ and $N_{prev}$ denote the set size and length of each constituent sequence, respectively, of the QCSS, corresponding to the optimality factor $\rho_{\min\_prev}$.
Our construction is different from \cite{Li19_3} and \cite{Li19_4} in following ways
\begin{enumerate}
	\item Our proposed construction can generate asymptotically optimal QCSS with more flexible parameters which are not covered by the results proposed in \cite{Li19_3} and \cite{Li19_4}. For example, the asymptotically optimal and near-optimal QCSS, given in Table \ref{tab_composite} and Table \ref{new_table_for_3}, respectively, can only be constructed by our construction till date.
	\item The constructions of asymptotically optimal QCSSs reported in \cite{Li19_3} are based on the construction of low correlation CSS, whereas our construction is based on the combination of multiple sets of CCCs.
	\item To have an unbiased comparison, we have also compared the cases when $N$ is a power of prime. In these cases, our set size is less than the previous results reported in \cite{Li19_3}. However, we obtain almost same optimality factor to that of the QCSS generated by the previous result reported in \cite{Li19_3}, for larger values of $N$. In Table \ref{tab_prime_power} we have listed down some cases, as example.
\end{enumerate}

\begin{table}[h!]
	\caption{The comparison when $N=p^2$, $p$ is prime.\label{tab_prime_power}}
	\small
	\renewcommand{\arraystretch}{1.3}
	\resizebox{\textwidth}{!}{
		\begin{tabular}{|c|c|c|c|c|c|c|c|}
			\hline
			Alphabet & $M$ & $N$& $N_{prev}$& $K$  &$K_{prev}$& $\rho$ &  $\rho_{\min\_prev}$   \\
			\hline
			$\mathbb{Z}_{11*11}$ & 121 & 121 &120& 1210 &14641& 1.2551 &  1.0614 \\
			\hline
			$\mathbb{Z}_{13*13}$& 169 & 169&168 & 2028  & 28561& 1.2247  & 1.0507 \\
			\hline
			$\mathbb{Z}_{17*17}$ & 289 & 289&280& 4624 &83521 & 1.1857 & 1.0376 \\
			\hline
			$\mathbb{Z}_{19*19}$ & 361 & 361&360& 6498 &130321& 1.1722 & 1.0333 \\
			\hline
			$\mathbb{Z}_{23*23}$ & 529 & 529&520& 11638  &279841& 1.1518 & 1.0271 \\
			\hline
			$\mathbb{Z}_{29*29}$ & 841 & 841&840&  23548  &707281& 1.1310 & 1.0211 \\
			\hline
			$\mathbb{Z}_{31*31}$ & 961 & 961&960& 28830  &923521& 1.1257 & 1.0197 \\
			\hline
			$\mathbb{Z}_{37*37}$ & 1369 & 1369&1360& 49284 &1874161& 1.1128 & 1.0164 \\
			\hline
			$\mathbb{Z}_{41*41}$ & 1681 & 1681&1680& 67240  &2825761& 1.1061 & 1.0147\\
			\hline
			$\mathbb{Z}_{43*43}$ & 1849& 1849&1848& 77658  &3418801& 1.1031 &1.0140\\
			\hline
			$\mathbb{Z}_{47*47}$ & 2209 & 2209&2208& 101614 &4879681 & 1.0978 & 1.0127\\
			\hline
	\end{tabular}}
\end{table}

\section{Concluding Remarks}
In this paper, we have presented a construction of QCSSs with new flexible parameters. We have relaxed the parameter $N\geq 5$ to be any odd integer. We first proposed a new permutation on $\mathbb{Z}_N$ and used that to construct $(N,N,N)$- CCCs, which lead to new sets of $(N(p_0-1),N,N,N)$- QCSS, where $p_0$ is the minimum prime factor of $N$. The proposed QCSSs are asymptotically optimal with respect to the correlation bound in \cite{zilong14_1}. We also consider the cases when $N$ is an odd integer having minimum prime factor $p_0=3$. In this case we have proposed $(2N,N,N,N)$- QCSS, which are near-optimal with respect to the Welch bound in \cite{welch}. Finally, we compare our proposed work with the previous works and shown that the parameters of the proposed QCSS are much more flexible as compared to the parameters of the QCSSs, reported in the previous constructions. An interesting future work will be to design optimal QCSSs over $\mathbb{Z}_N$, when $N$ is even (but not a power of prime) or a composite integer having minimum prime factor $3$.

\section*{Appendix A\\Proof of Theorem 1}

	Let us prove the first part. Let $\mathcal{C}^{\left(k, m_{1}\right)}, \mathcal{C}^{\left(k, m_{2}\right)} \in \mathfrak{C}^{k},$ where $1\leq k <p_0$, $0 \leq m_{1}, m_{2} \leq N-1$ and $f_{s}^{(k,m)}(t)$ follows (\ref{eq_new_const2}). Then the aperiodic correlation of  $\mathcal{C}^{\left(k, m_{1}\right)}$ and $\mathcal{C}^{\left(k, m_{2}\right)}$ is

\begin{equation}
\begin{split}
\sum_{s=0}^{N-1} & \tilde{R}_{C_{s}^{\left(k, m_{1}\right)}, C_{s}^{\left(k, m_{2}\right)}}(\tau)\\& =\sum_{s=0}^{N-1} \sum_{t=0}^{N-1-\tau} C_{s, t}^{\left(k, m_{1}\right)} \cdot\left(C_{s, t+\tau}^{\left(k, m_{2}\right)}\right)^{*} \\
& =\sum_{s=0}^{N-1} \sum_{t=0}^{N-1-\tau} \omega_{N}^{f_{s}^{\left(k, m_{1}\right)}(t)} \cdot {\omega_{N}}^{-f_{s}^{\left(k, m_{2}\right)}(t+\tau)}\\
& =\sum_{s=0}^{N-1} \sum_{t=0}^{N-1-\tau} \omega_{N}^{k s(\pi(t)-\pi(t+\tau))+t\left(m_{1}-m_{2}\right)-m_{2} \tau}.
\end{split}
\end{equation}

We have the following four cases to consider.

Case 1: When $m_1=m_2,\tau=0$, we have
\begin{equation}
\sum_{s=0}^{N-1} \tilde{R}_{C_{s}^{\left(k, m_{1}\right)}, C_{s}^{\left(k, m_{2}\right)}}(0)=N^2.
\end{equation}

Case 2: When $m_1=m_2,1\leq\tau\leq N-1$,
\begin{equation}\label{eq22}
\begin{split}
\sum_{s=0}^{N-1}& \tilde{R}_{C_{s}^{\left(k, m_{1}\right)},{ C_{s}^{\left(k, m_{2}\right)}}}(\tau) \\& =\sum_{s=0}^{N-1} \sum_{t=0}^{N-1-\tau} \omega_{N}^{-m_{2} \tau} \cdot \omega_{N}^{k s(\pi(t)-\pi(t+\tau))} \\
& =\omega_{N}^{-m_{2} \tau} \cdot \sum_{t=0}^{N-1-\tau} \sum_{s=0}^{N-1} \omega_{N}^{ks(\pi(t)-\pi(t+\tau))}=0.
\end{split}
\end{equation}
Since $\pi(t)$ is a permutation on $\mathbb{Z}_N$, therefore for $\tau \neq 0$, $\pi(t)\neq \pi(t+\tau)$. Also $N\nmid k(\pi(t)-\pi(t+\tau))$ when $1\leq k <p_0$. Therefore (\ref{eq22}) holds.\\

Case 3: When $m_1\neq m_2,\tau=0$,
\begin{equation}
\sum_{s=0}^{N-1} \tilde{R}_{C_{s}^{\left(k, m_{1}\right)},{C_{s}^{\left(k, m_{2}\right)}}}(0)=\sum_{s=0}^{N-1} \sum_{t=0}^{N-1} \omega_{N}^{t \left(m_{1}-m_{2}\right)}=0.
\end{equation}

Case 4: When $m_1\neq m_2,1\leq\tau\leq N-1$,
\begin{equation}\label{neq30}
\begin{split}
\sum_{s=0}^{N-1} &\tilde{R}_{C_{s}^{\left(k, m_{1}\right)},{ C_{s}^{\left(k, m_{2}\right)}}}(\tau)\\&=\sum_{t=0}^{N-1-\tau} \omega_{N}^{t \cdot\left(m_{1}-m_{2}\right)-m_{2} \tau} \cdot \sum_{s=0}^{N-1}\omega_{N}^{ks(\pi(t)-\pi(t+\tau))}=0.
\end{split}
\end{equation}
Since $\pi(t)$ is a permutation on $\mathbb{Z}_N$, therefore for $\tau \neq 0$, $\pi(t)\neq \pi(t+\tau)$. Also $N\nmid k(\pi(t)-\pi(t+\tau))$ when $1\leq k <p_0$. Therefore (\ref{neq30}) holds.

The four cases above illustrate that each sequence set $\mathfrak{C}^{k}$ is a $(N,N,N)$-CCC.

We now prove the second part.
Let us consider $\mathcal{C}^{\left(k_{1}, m_{1}\right)} \in \mathfrak{C}^{k_1}$ and $\mathcal{C}^{\left(k_{2}, m_{2}\right)} \in \mathfrak{C}^{k_2}$. Then the aperiodic correlation of $\mathcal{C}^{\left(k_{1}, m_{1}\right)} $ and  $\mathcal{C}^{(k_{2}, m_{2})}$ is
\begin{equation}
\begin{split}
\sum_{s=0}^{N-1} &\tilde{R}_{C_{s}^{\left(k_{1}, m_{1}\right)},{C_{s}^{\left(k_{2}, m_{2}\right)}}}(\tau) \\& =\sum_{s=0}^{N-1} \sum_{t=0}^{N-1-\tau} \omega_{N}^{f_{s}^{\left(k_{1}, m_{1}\right)}(t)} \cdot \omega_{N}^{-f_{s}^{\left(k_{2}, m_{2}\right)}(t+\tau)} \\
& =\sum_{s=0}^{N-1} \sum_{t=0}^{N-1-\tau} \omega_{N}^{t\left(m_{1}-m_{2}\right)-m_{2} \tau+s\left(k_{1} \pi(t+\tau)-k_{2} \pi(t)\right)}\\
& =\sum_{t=0}^{N-1-\tau} \omega_{N}^{t\left(m_{1}-m_{2}\right)-m_{2} \tau} \cdot \sum_{s=0}^{N-1} \omega_{N}^{s\left(k_{1} \pi(t+\tau)-k_{2} \pi(t)\right)}.
\end{split}
\end{equation}

Recall that permutation $\pi$ satisfies Lemma \ref{new_per_lem}. Therefore, $k_{1} \pi(t)-k_{2} \pi(t+\tau)\equiv0\pmod N$ for any $0 \leq \tau \leq N-1, k_1 \neq k_2$ has only one solution. Therefore, for $0\leq t^{\prime}\leq N-1$ and  $k_{1} \pi(t')-k_{2} \pi(t'+\tau)=0 \pmod N$. If $N-1\geq t^{\prime} \geq N-\tau$, then $\sum\limits_{s=0}^{N-1} \tilde{R}_{C_{s}^{\left(k_{1}, m_{1}\right)}, C_{s}^{\left(k_{2}, m_{2}\right)}}(\tau)=0$ due to $\sum\limits_{s=0}^{N-1} \omega_{N}^{s\left(k_{2} \pi(t+\tau)-k_{1} \pi(t)\right)}=0$. For $0\leq t'\leq N-1-\tau$, we have
\begin{equation}
\begin{split}
\sum_{s=0}^{N-1}& \tilde{R}_{C_{s}^{\left(k_{1}, m_{1}\right)},{C_{s}^{\left(k_{2}, m_{2}\right)}}}(\tau)
\\& =\omega_{N}^{-m_{2} \tau} \cdot[\omega_{N}^{(m_{1}-m_{2})\cdot t' }\cdot  N+ \nonumber \\
& \sum_{\substack{0\leq t\leq N-1-\tau, \\ t \neq t'}} \omega_{N}^{\left(m_{1}-m_{2}\right) t}  \sum_{0\leq s\leq N-1} \omega_{N}^{(k_{2} \pi(t+\tau)-k_{1} \pi(t)) \cdot s}] \nonumber\\
& =\omega_{N}^{-m_{2} \tau+\left(m_{1}-m_{2}\right) t^{\prime}} \cdot N.
\end{split}
\end{equation}
Hence, $|\sum\limits_{s=0}^{N-1} \tilde{R}_{C_{s}^{\left(k_{1}, m_{1}\right)}, {C_{s}^{\left(k_{2}, m_{2}\right)} }}(\tau)| \leq N$ for all $0\leq\tau \leq N-1, k_1\neq k_2$ and $1\leq m_1, m_2\leq N-1.$

Hence, the theorem follows.

\end{document}